\documentclass[a4paper,11pt,notitlepage]{article}
\usepackage{multirow,url,color}
\usepackage[utf8x]{inputenc}

\title{\textbf{Effects of irradiation on Triple and Single Junction InGaP/GaAs/Ge solar cells}}
\author{Roberta Campesato$^{1}$, Carsten Baur$^{5}$, Mariacristina Casale$^{1}$, Massimo Gervasi$^{2,3}$,\\ Enos Gombia$^{4}$, Erminio Greco$^{1}$, Aldo Kingma$^{4}$, Pier Giorgio  Rancoita$^{2}$,\\ Davide Rozza$^{2,3}$, Mauro Tacconi$^{2,3}$.}
\date{}

\usepackage[utf8x]{inputenc}
\usepackage{amsmath}
\usepackage{amsfonts}
\usepackage{amssymb}
\usepackage{graphicx}
\usepackage[top=2.3cm,bottom=2.3cm,left=2.3cm,right=2.3cm]{geometry}

\begin{document}
\maketitle
\begin{center}
$^1$ \textit{CESI, via Rubattino 54, I-20134 Milan, Italy}\\
$^2$ \textit{INFN Sezione di Milano Bicocca, I-20126 Milan, Italy}\\
$^3$ \textit{Universit\`a di Milano Bicocca, I-20126 Milan, Italy} \\
$^4$ \textit{IMEM-CNR Institute, Parco Area delle Scienze 37/A, 43124 Parma, Italy}\\
$^5$ \textit{ESA/ ESTEC, Keplerlaan 1, 2201 AZ Noordwijk, The Netherlands}\\
\vspace{0.5cm}
To appear in the Proceedings of the 35th European PV Solar Energy Conference, \\Brussels, 24-28 September 2018.
\end{center}

\begin{abstract}
The investigation of the degradation effects on triple-junction (TJ) solar cells, operating in space environment, is of primary importance in view of future space missions towards harsh radiation orbits (e.g. MEO with high particle irradiation intensity) and for the new spacecraft based on electrical propulsion.\\
In the present work, we report the experimental results obtained from the irradiation test campaign carried out both on (InGaP/GaAs/Ge) TJ solar cells and on single junction (SJ) isotype sub-cells (i.e. InGaP top cells, GaAs middle cells and Ge bottom cells).\\
In particular, the electrical performances degradation of the irradiated samples was analyzed by means of the Displacement Damage Dose (DDD) method based on the NIEL (Non Ionizing Energy Loss) scaling hypothesis, with DDD doses computed by the SR-NIEL approach.\\
By plotting the remaining factors as a function of the DDD, a single degradation curve for each sample type and for each electrical parameter has been found. The collapse of the remaining factors on a single curve has been achieved by selecting suitable displacement threshold energies, $E_d$, for NIEL calculation.\\
Moreover, DLTS (Deep level Transient Spectroscopy) technique was used to better understand the nature of the defects produced by irradiation. DLTS measurements reveal a good correlation between defects introduced by irradiation inside the GaAs sub-cell and the calculated Displacement Damage Doses.
\end{abstract}

\section{Introduction}
The prediction of the solar cell degradation due to particle irradiation is one of the key elements to properly size a solar array for a given mission.\\
It is known that the degradation of the solar cells electrical performances is a function of the energy, the fluence and the type of the incident particles (electron or proton).\\
Since the space environment can be approximated by an omnidirectional spectrum of particles incident on solar cells, a method is needed to predict the on-orbit solar cell performance.\\
There are two methodologies currently available to perform on-orbit solar cell performance predictions. The first one is the Equivalent Fluence Method developed by JPL (Jet Propulsion Laboratory) \cite{Anspaugh} and the second one is the Displacement Damage Dose (DDD) Method developed by NRL (Naval Research Laboratory) \cite{Messengers}.\\
The JPL approach is the first implemented and used by the space actors. To employ this method, it is necessary to determine the degradation curves for each electrical parameter (e.g. $V_{oc}$, $I_{sc}$, $P_{max}$ etc.) over a wide range of energies both for protons and electrons. Each degradation curve is obtained by plotting the remaining factor (the ratio of the end-of-life EOL value to the beginning-of-life value) of a given electrical parameter as a function of the fluence. Therefore, this method requires a significant number of irradiation tests with electrons and protons with various energies.\\
NRL approach is based on the hypothesis that the permanent displacement damage produced by nonionizing events is the main phenomenon that degrades the device electrical performance. With this approach a single characteristic degradation curve, for each electrical parameter, can be obtained. This occurs if the remaining factors are plotted against the displacement damage dose, rather than fluence. The displacement damage dose (DDD) is determined from a calculation of the nonionizing energy loss (NIEL); for electrons, protons, ions and neutrons passing through an absorber (elemental or compound) the NIEL values can be obtained by means SR-NIEL framework approach (\cite{Boschini} and Chapters 2, 7 and 11 in \cite{Leroy}). The main advantage of the latter method is that it requires less irradiation tests (e.g. only 1 energy for proton and 2 energies for electrons) to predict the EOL behavior of solar cells.\\
In this paper, the results of electron and proton irradiation tests, performed on triple junction solar cells and related isotype sub-cells manufactured by CESI, are presented.  In particular, the radiation hardness analysis has been conducted by the DDD method.\\
The dependence of the NIEL calculation on the threshold energies for displacement, $E_{d}$, is the key point to obtain a single degradation curve as a function of the DDD. The value of the energy threshold has been chosen to make collapse the experimental data in a single degradation curve for each solar cell electrical parameter.\\
In parallel to this analysis, DLTS (Deep level Transient Spectroscopy) technique has been used on InGaP (top sub-cell) and GaAs (middle sub-cell) samples manufactured as diodes and irradiated together with the solar cells. The diodes were manufactured with exactly the same epitaxial structure of the sub-cells composing the triple junction device.\\
From the DLTS measurements, it turned out that the concentration of defects introduced by irradiation on GaAs samples is a linear function of the DDD. In particular, the expected linear behavior is obtained if the same value $E_d$ used to make collapse the degradation curves of the GaAs sub-cell, is selected.

\section{Irradiation test campaign on TJ solar cells, component cells and diodes}
\begin{figure}[h!]
\centering
\includegraphics[width=0.7\textwidth]{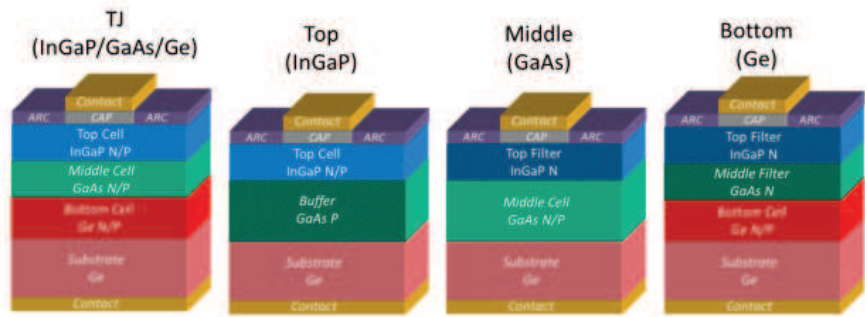}
\caption{Scheme of triple junction (TJ) and single junction (SJ) isotype sub-cells (Top, Middle and Bottom).}
\label{fig:schema}
\end{figure}

\begin{figure}[h!]
\centering
\includegraphics[width=0.7\textwidth]{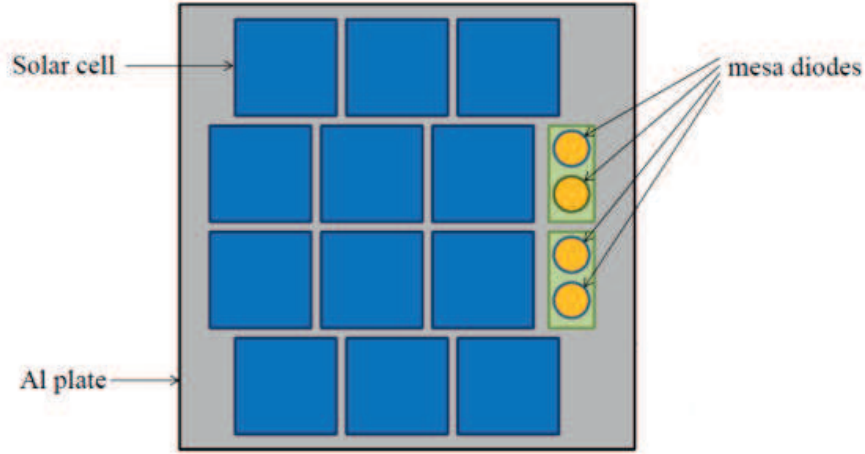}
\caption{Arrangement of solar cells and diodes on Al plate for proton irradiation.}
\label{fig:Arrangement}
\end{figure}

\begin{figure}[h!]
\centering
\includegraphics[width=0.5\textwidth]{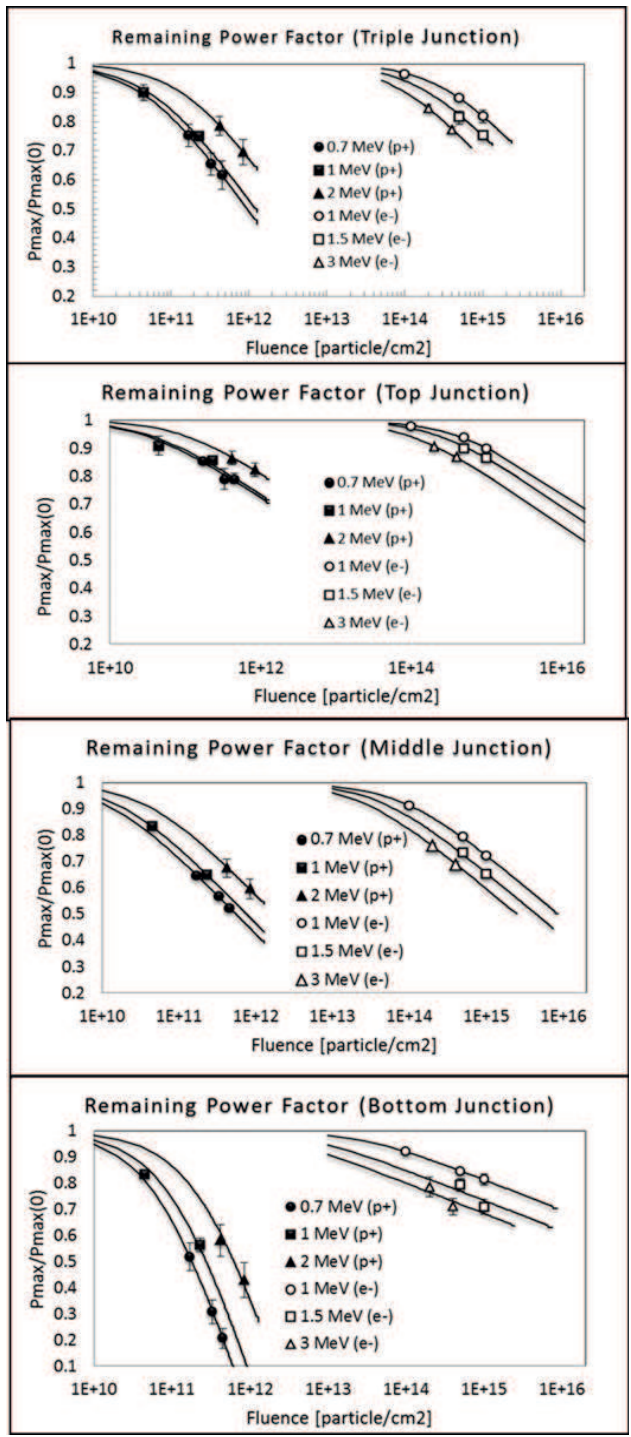}
\caption{$P_{max}$ degradation curves of TJ, top, middle and bottom solar cells as a function of the particle fluence.}
\label{fig:Pmaxdeg}
\end{figure}

InGaP/InGaAs/Ge TJ solar cells and related component cells with AM0 efficiency class 30$\%$  (CTJ30) \cite{Gori} have been manufactured as $2\times2$ cm$^2$ solar cells.\\
The basic structure of the solar cells is reported in Figure \ref{fig:schema}.
The TJ solar cell is composed by a Ge bottom junction obtained by diffusion into the germanium P-type substrate, a middle junction of (In)GaAs, whose energy gap is around 1.38 eV and a top junction of InGaP with an energy gap of 1.85 eV. Component cells are single-junction cells which are an electrical and optical representation of the sub-cells inside the TJ cell. Therefore, to manufacture them, special attention was put to reproduce the optical thicknesses of all the upper layers present in the TJ structure.\\
Top and middle sub-cells were also manufactured as diodes with 0.5 mm diameter for DLTS analysis before and after irradiation. In Figure \ref{fig:Arrangement} is shown how the solar cells and the diodes were arranged for irradiation tests.\\
In Table \ref{Tab1} the energy-fluence pairs used for protons and electrons irradiation tests are reported. For each condition, the following ($2\times2$ cm$^2$) cells were irradiated: 3 TJ, 3 Top, 3 Middle and 3 Bottom. InGaP and GaAs diodes, for DLTS analysis, were irradiated together with the solar cells.\\
\begin{table}[h!]
\centering
\caption{Electron and proton irradiation test campaign dedicated to establish a degradation data set.}
\begin{tabular}{|c|c|c|c|}
\hline
\multicolumn{2}{|c|}{Protons} & \multicolumn{2}{|c|}{Electrons} \\
\hline
\multicolumn{1}{|c|}{\textbf{Energy [MeV]}} & 
\multicolumn{1}{|c|}{\textbf{Fluence [p+/cm$^2$]}} & 
\multicolumn{1}{|c|}{\textbf{Energy [MeV]}} & 
\multicolumn{1}{|c|}{\textbf{Fluence [e-/cm$^2$]}}\\
\hline
\multicolumn{1}{|c|}{\multirow{1}{*}{\begin{tabular}{c}0.7\\\end{tabular}}} & 
\multicolumn{1}{|c|}{1.7E+11} & 
\multicolumn{1}{|c|}{\multirow{1}{*}{\begin{tabular}{c}1\\\end{tabular}}} & 
\multicolumn{1}{|c|}{1.0E+14} \\
\multicolumn{1}{|c|}{} & 
\multicolumn{1}{|c|}{3.3E+11} & 
\multicolumn{1}{|c|}{} & 
\multicolumn{1}{|c|}{5.0E+14} \\
\multicolumn{1}{|c|}{} & 
\multicolumn{1}{|c|}{4.5E+11} & 
\multicolumn{1}{|c|}{} & 
\multicolumn{1}{|c|}{1.0E+15} \\
\hline
\multicolumn{1}{|c}{\multirow{1}{*}{\begin{tabular}{c}1\\\end{tabular}}} & 
\multicolumn{1}{|c}{4.5E+10} & 
\multicolumn{1}{|c}{\multirow{1}{*}{\begin{tabular}{c}1.5\\\end{tabular}}} & 
\multicolumn{1}{|c|}{5.0E+14} \\
\multicolumn{1}{|c}{} & 
\multicolumn{1}{|c}{2.3E+11} & 
\multicolumn{1}{|c}{} & 
\multicolumn{1}{|c|}{1.0E+15} \\
\hline
\multicolumn{1}{|c}{\multirow{1}{*}{\begin{tabular}{c}2\\\end{tabular}}} & 
\multicolumn{1}{|c}{4.2E+11} & 
\multicolumn{1}{|c}{\multirow{1}{*}{\begin{tabular}{c}3\\\end{tabular}}} & 
\multicolumn{1}{|c|}{2.0E+14} \\
\multicolumn{1}{|c}{} & 
\multicolumn{1}{|c}{8.4E+11} & 
\multicolumn{1}{|c}{} & 
\multicolumn{1}{|c|}{4.0E+14} \\
\hline
\end{tabular}\label{Tab1}
\end{table}
To calculate the remaining factors of each electrical parameter, all the solar cells have been measured both in BOL and EOL conditions. In particular, two measurements have been performed after the irradiation tests: after self-annealing (i.e. solar cells kept in dry box for 1 month before measurement) and after photon and thermal annealing (i.e. 8 hours at 1 AM0 illumination followed by 48 h at 60$ ^{\circ}$ C according to  ECSS standards).\\
In Figure \ref{fig:Pmaxdeg} the $P_{max}$ degradation curves after self-annealing of TJ, top, middle and bottom cells are shown.  The following semi-empirical equation (\ref{Eq1}) is used to fit the experimental $P_{max}$ Remaining Factors (for each particle/energy separately):
\begin{equation}\label{Eq1}
RPF= \frac{P_{max} \left( n,E, \phi  \right) }{P_{max} \left( BOL \right) }=1-C\cdot\log_{10} \left( 1+\frac{ \phi }{ \phi _{crit} \left( n,E \right) } \right)
\end{equation}
with $\Phi$ being the particle fluence and with $C$, $\Phi_{crit}$ (critical fluence) being fitting parameters. The parameters n, E refer to the particle type and the particle energy indicating that the degradation and the critical fluence are particle and energy dependent. Equivalent curves can be plotted for each electrical parameter (i.e. $V_{oc}$, $I_{sc}$).\\
Analysing the curves in Figure \ref{fig:Pmaxdeg}, it turned out that the samples irradiated with electrons degrade as expected, whereas a higher spread in the remaining power factors was observed for the samples irradiated with protons, especially for the TJ and Ge bottom cells. It can also be noticed that Ge bottom junction is high resistant against electrons but it shows a bad radiation resistance against proton irradiation, especially if low energy protons are concerned \cite{Baur}. Moreover, top junction (InGaP) confirms its strong behaviour against irradiation. The analysis of remaining factors against DDD is reported in the next chapter.\\

\section{NIEL analysis}
In this section, the degradation of the photovoltaic parameters of TJ solar cells and single junction sub-cells is investigated using the DDD method.\\
According to this approach, the remaining factors of each electrical parameter are plotted as a function of the displacement damage dose (DDD).  This quantity, expressed in units of MeV/g, can be obtained from equation (\ref{Eq2}):
\begin{equation}\label{Eq2}
DDD=\Phi\frac{dE_{de}}{d\chi}
\end{equation}
where $\Phi$  is the fluence in cm$^{-2}$ of traversing particles (electrons or protons) and $\frac{dE_{de}}{d\chi}$ is the so-called non-ionizing energy loss NIEL which expresses the amount of energy deposited by an incident particle passing through a material and resulting in displacement processes. This quantity, expressed in units of MeVcm$^2$/g, can be calculated by means of the SR (Screened Relativist) treatment (Eq. \ref{Eq3}) ([3,4]):
\begin{equation}\label{Eq3}
\frac{dE_{de}}{d \chi }=\frac{N}{A} \int _{E_{d}}^{E_{R}^{max}}E_{R}L \left( E_{R} \right) \frac{d \sigma  \left( E,E_{R} \right) }{dE_{R}}dE_{R}
\end{equation}
where $ \chi $  is the absorber thickness in g/cm$^2$; $N$ is the Avogadro constant, $A$ is the atomic weight of the medium; $E$ is the kinetic energy of the incoming particle; $E_R$ and $E_R^{max}$ are the recoil kinetic energy and the maximum energy transferred to the recoil nucleus respectively; $L(E_R)$ is the Lindhard partition function; $d\sigma(E,E_R)/dE_R$ is the differential cross section for elastic Coulomb scattering for electrons or protons on nuclei; $E_d$ is the so-called displacement threshold energy, i.e. the minimum energy necessary to permanently displace an atom from its lattice position.\par
To perform the calculations, the energy of the incoming particles E was modified considering the energy losses along the path inside the non-active absorbing layers. In particular, it turns out that there is no relevant kinetic energy variation for electrons whereas the losses are not negligible for protons. The actual kinetic energy E, in each junction, was estimated by means of SRIM \cite{Ziegler}.\par
Moreover, the displacement threshold energy $E_d$ has been suitably chosen to make collapse the remaining factors into a single degradation curve for each type of cell. The degradation curves can be fitted with a function analogous to equation (\ref{Eq1}) where the fluence is replaced with the calculated DDD (Eq.\ref{Eq4}):
\begin{equation}\label{Eq4}
\frac{P_{max} \left( EOL \right) }{P_{max} \left( BOL \right) }=A-C \cdot \log_{10} \left[ 1+\frac{DDD \left( E_{d} \right) }{DDD_{x}} \right]
\end{equation}
being A, C and DDD$_x$ fitting parameters.\par
\begin{figure}[h!]
\centering
\includegraphics[width=0.5\textwidth]{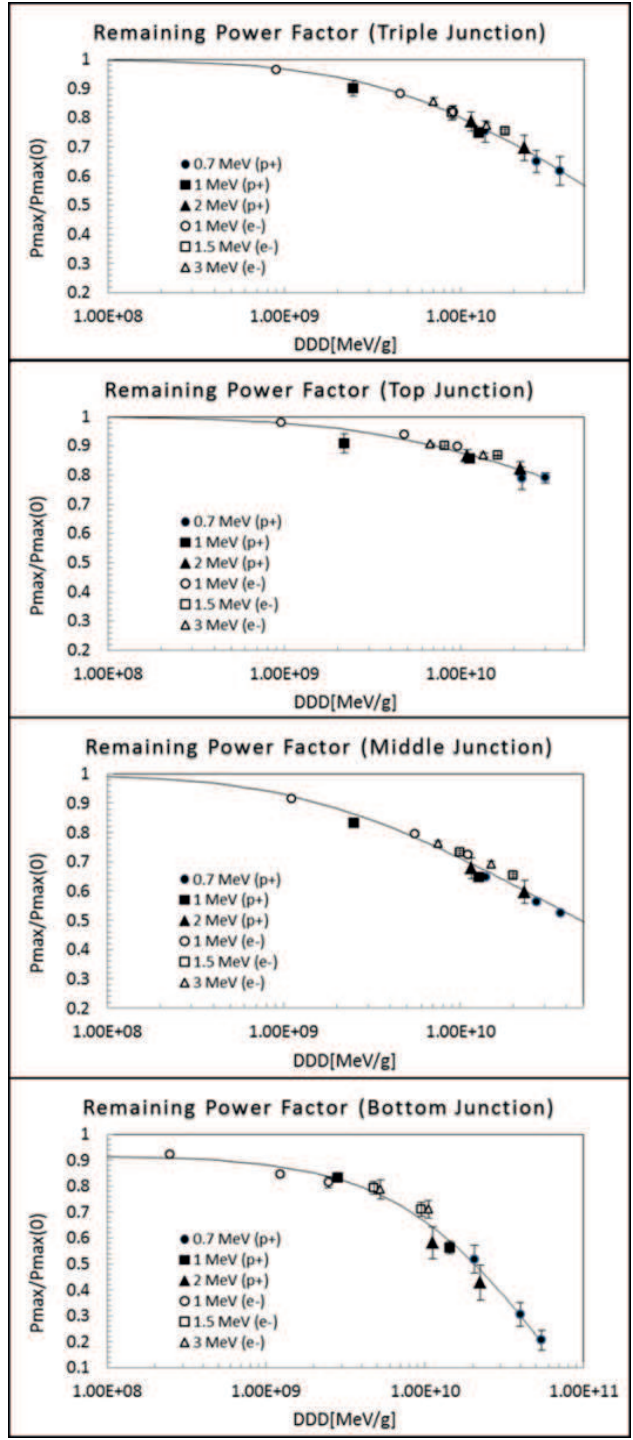}
\caption{$P_{max}$ degradation curves of TJ, top, middle and bottom solar cells as a function of the DDD.}
\label{fig:4}
\end{figure}

The $P_{max}$ degradation curves as a function of the DDD for TJ, top, middle and bottom cells are shown in figure \ref{fig:4}.\par
The optimal fit for each cell type, was obtained using the $E_d$ values reported in Table \ref{Tab2}.\\
In this work, the TJ solar cell was approximated to a GaAs cell single junction cell since the middle cell is the one that mainly affects the overall TJ performances.
\begin{table}[h!]
\centering
\caption{Optimum $E_d$ values found to obtain the best-fit curves shown in Fig. \ref{fig:4}.}
\begin{tabular}{|c|c|}
\hline
\multicolumn{1}{|c}{\textbf{Cell type}} & 
\multicolumn{1}{|c|}{\textbf{$E_d$}} \\
\hline
\multicolumn{1}{|c}{\multirow{1}{*}{\begin{tabular}{c}InGaP\\\end{tabular}}} & 
\multicolumn{1}{|c|}{$E_d$(In) $ \approx $  43eV} \\
\multicolumn{1}{|c}{} & 
\multicolumn{1}{|c|}{$E_d$(Ga) $ \approx $  21eV} \\
\multicolumn{1}{|c}{} & 
\multicolumn{1}{|c|}{$E_d$(P) $ \approx $  21eV} \\
\hline
\multicolumn{1}{|c}{\multirow{1}{*}{\begin{tabular}{c}GaAs\\\end{tabular}}} & 
\multicolumn{1}{|c|}{$E_d$(Ga) $ \approx $  21eV} \\
\multicolumn{1}{|c}{} & 
\multicolumn{1}{|c|}{$E_d$(As) $ \approx $  21eV} \\
\hline
\multicolumn{1}{|c}{Ge} & 
\multicolumn{1}{|c|}{$E_d$(Ge) $ \approx $  40eV} \\
\hline
\multicolumn{1}{|c}{\multirow{1}{*}{\begin{tabular}{c}TJ (GaAs)\\\end{tabular}}} & 
\multicolumn{1}{|c|}{$E_d$(Ga) $ \approx $  24eV} \\
\multicolumn{1}{|c}{} & 
\multicolumn{1}{|c|}{$E_d$(As) $ \approx $  24eV} \\
\hline
\end{tabular}\label{Tab2}
\end{table}
A single degradation curve has been obtained also for $V_{oc}$ and $I_{sc}$ remaining factors using the same $E_{d}$ values reported in Table \ref{Tab2}.\\
In the next chapter, we will see how the $E_{d}$ value for GaAs has been confirmed from the DLTS analysis performed on GaAs middle junction.

\section{DLTS analysis}
\begin{figure}[h!]
\centering
\includegraphics[width=0.5\textwidth]{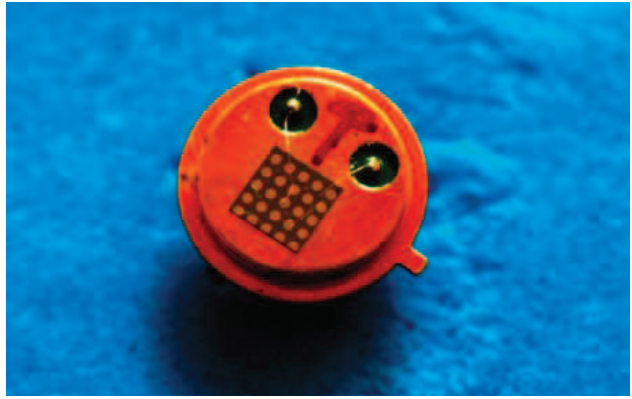}
\caption{0.5 mm dia diodes for DLTS and irradiation analysis.}
\label{fig:5}
\end{figure}

In order to perform DLTS (Deep Level Transient Spectroscopy) investigations of deep levels induced by electron and proton irradiation, diodes of 0.5 mm in diameter have been prepared using the same epitaxial structure of top and middle sub-cells. The samples were manufactured by means of optical lithography, evaporation of metal contacts and mesa etch to remove the edge defects. Figure \ref{fig:5} shows a picture of a diodes rack used for DLTS analysis.\\
Concerning the InGaP diode (Top sub-cell), due to the relatively high doping concentration and the low reverse bias that can be applied to it, it was only possible to investigate regions near the junction interface dominated by high densities of interface states. These restrictions prevented a detailed study of bulk deep levels on both irradiated and non-irradiated samples.\\
For this reason, the DLTS study has been mainly focused on GaAs diodes (Middle sub-cell).\\
\begin{figure}[h!]
\centering
\includegraphics[width=0.65\textwidth]{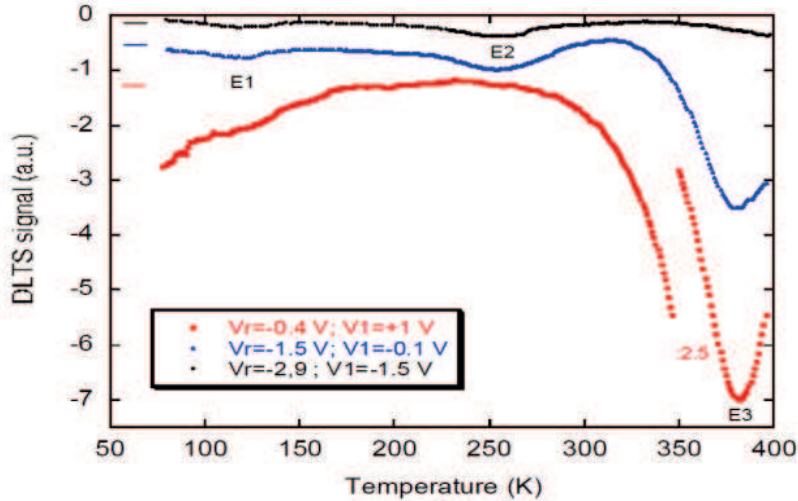}
\caption{DLTS spectra obtained on a middle junction irradiated with electrons (1 MeV at a fluence of $10^{15}$ cm$^{-2}$) using different reverse voltage $V_r$ at fixed pulse amplitude of 1.4 V. Emission rate =46 s$^{-1}$, pulse width=500 ms, V1=pulse voltage.}
\label{fig:6}
\end{figure}
Figure \ref{fig:6} shows the DLTS spectra of a GaAs diode irradiated with electrons\  using different reverse voltages $V_{r}$ (-0.4 V, -1.5 V, and -2.9 V) at fixed pulse amplitude (1.4 V).\\
With increasing reverse voltage $V_{r}$ the spectra show the characteristics of regions at deeper distance from the n+/p interface. The high temperature peak (activation energy = 0.7 eV) labelled E3, which is also present in the non-irradiated samples, is likely to correspond to a defect at the junction interface. For higher reverse biases the DLTS spectra show the presence of at least two levels, labelled E1 (0.21 eV) and E2 (0.45 eV).\\
These levels have been identified as majority carrier traps in the p-type bulk bases since majority carrier pulses have been used in the DLTS measurements. As a consequence their energy must be referred to the top of the valence band.\\
These defect states are not observed in non-irradiated samples and show a density which increases with the absorbed dose, hence they are attributed to irradiation induced defects.\\
\begin{figure}[h!]
\centering
\includegraphics[width=0.65\textwidth]{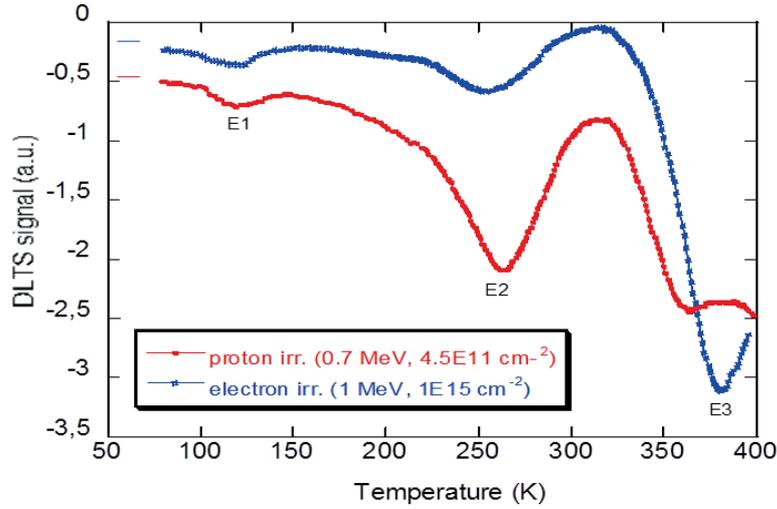}
\caption{Comparison of the DLTS spectra of two middle junctions irradiated by protons and electrons respectively. Emission rate=46 s$^{-1}$, pulse width=500$ \mu$s, reverse voltage $V_r$=-1.5 V, pulse voltage V1= -0.1 V}
\label{fig:7}
\end{figure}
In Figure \ref{fig:7} the DLTS spectrum of a middle cell diode irradiated with protons (energy 0.7 MeV and fluence $4.5\cdot10^{11}$cm$^{-2}$ corresponding to NIEL dose of $3\cdot10^{10}$ MeV g$^{-1}$) is compared to that of a middle cell diode irradiated with electrons (energy 1 MeV and fluence $1\cdot10^{15}$cm$^{-2}$ corresponding to NIEL dose of $1.1\cdot10^{10}$MeV g$^{-1}$).\\
From the analysis of the DLTS graphs obtained at different combinations of particle-energy, it can be observed that:
\begin{itemize}
\item The peak E1 is detected at all doses for samples irradiated with electrons whereas at only highest doses for those irradiated with protons;
\item The peak E2 is detected at all doses both for electron and proton irradiated samples.
\end{itemize}
From the DLTS peak height of the different irradiated diodes, it is possible to estimate the defects concentration as a function of the DDD.\\
\begin{figure}[h!]
\centering
\includegraphics[width=0.65\textwidth]{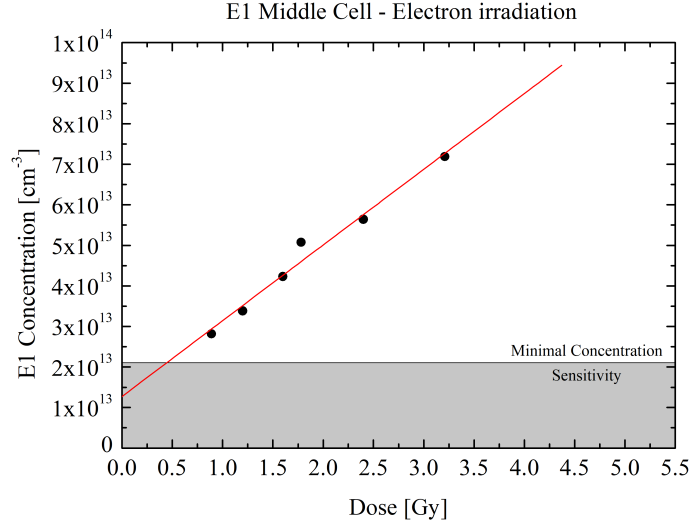}
\caption{Concentration of E1 traps induced by electron irradiation in middle sub cell as a function of Displacement Damage Dose (obtained with $E_d$ = 21 eV).}
\label{fig:8}
\end{figure}
\begin{figure}[h!]
\centering
\includegraphics[width=0.65\textwidth]{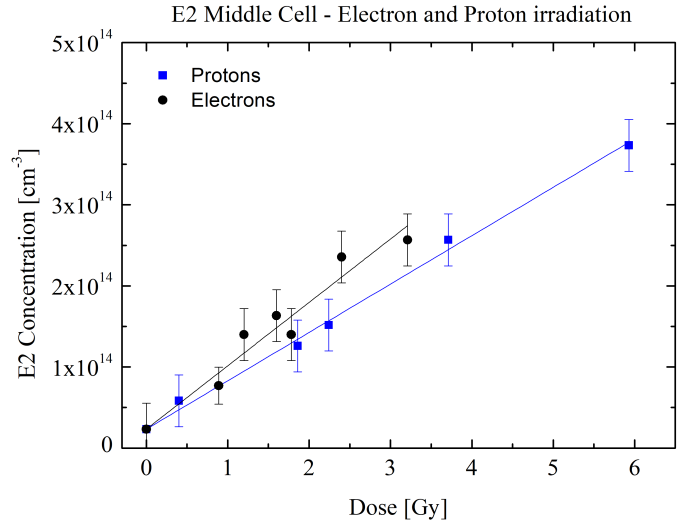}
\caption{Concentration of E2 traps induced by electron and proton irradiation in middle sub cell as a function of Displacement Damage Dose (obtained with $E_d$ = 21 eV).}
\label{fig:9}
\end{figure}
In particular, the concentration of E1 traps was correlated to the DDD of the GaAs diodes irradiated with electrons (Figure \ref{fig:8}) and the concentration of E2 traps was correlated to the DDD of the GaAs diodes irradiated both with electrons and protons (Figure \ref{fig:9}). In both cases, the expected linear behavior is obtained using the same threshold energies $E_{d}$, selected for the NIEL analysis of the previous paragraph (i.e. $E_{d}$ = 21 eV for GaAs). Concerning the peak E2, two independent linear fit with slightly different slopes have been found for protons and electrons.\\
The good correlation found between the DLTS measurements and the NIEL analysis can be appreciated considering the Trap Introduction Rate (TIR) of the defect states E1 and E2 as a function of the energy E of the incident particle (Eq. \ref{Eq5}):
\begin{equation}\label{Eq5}
TIR \left( E \right) = \frac{\textrm{Trap concentration}  \left[\textrm{cm}^{-3} \right] }{ \Phi  \left( E \right)  \left[\textrm{cm}^{-2} \right] }
\end{equation}
It turns out that the trap introduction rate TIR(E) is proportional to NIEL(E) (in units of MeV cm$^{2}$/g) being the proportionality constant given by the slope of the lines in Figure 8 and in Figure 9 (Traps Concentration vs DDD).\\
The conversion factors between trap introduction rate and NIEL are reported in Table \ref{Tab3}.
\begin{table}[h!]
\centering
\caption{Conversion factors between Trap Introduction Rate TIR(E) and NIEL.}
\begin{tabular}{|c|c|c|}
\hline
\multicolumn{1}{|c}{\textbf{Traps}} & 
\multicolumn{1}{|c}{\textbf{Particle}} & 
\multicolumn{1}{|c|}{\textbf{TIR/NIEL(g/MeV/cm$^{3}$)}} \\
\hline
\multicolumn{1}{|c}{E1} & 
\multicolumn{1}{|c}{electron} & 
\multicolumn{1}{|c|}{$3.0\cdot10^{3}$} \\
\hline
\multicolumn{1}{|c}{E2} & 
\multicolumn{1}{|c}{electron} & 
\multicolumn{1}{|c|}{$1.2\cdot10^{4}$} \\
\hline
\multicolumn{1}{|c}{E2} & 
\multicolumn{1}{|c}{proton} & 
\multicolumn{1}{|c|}{$9.3\cdot10^{3}$} \\
\hline
\end{tabular}\label{Tab3}
\end{table}
\begin{figure}[h!]
\centering
\includegraphics[width=0.5\textwidth]{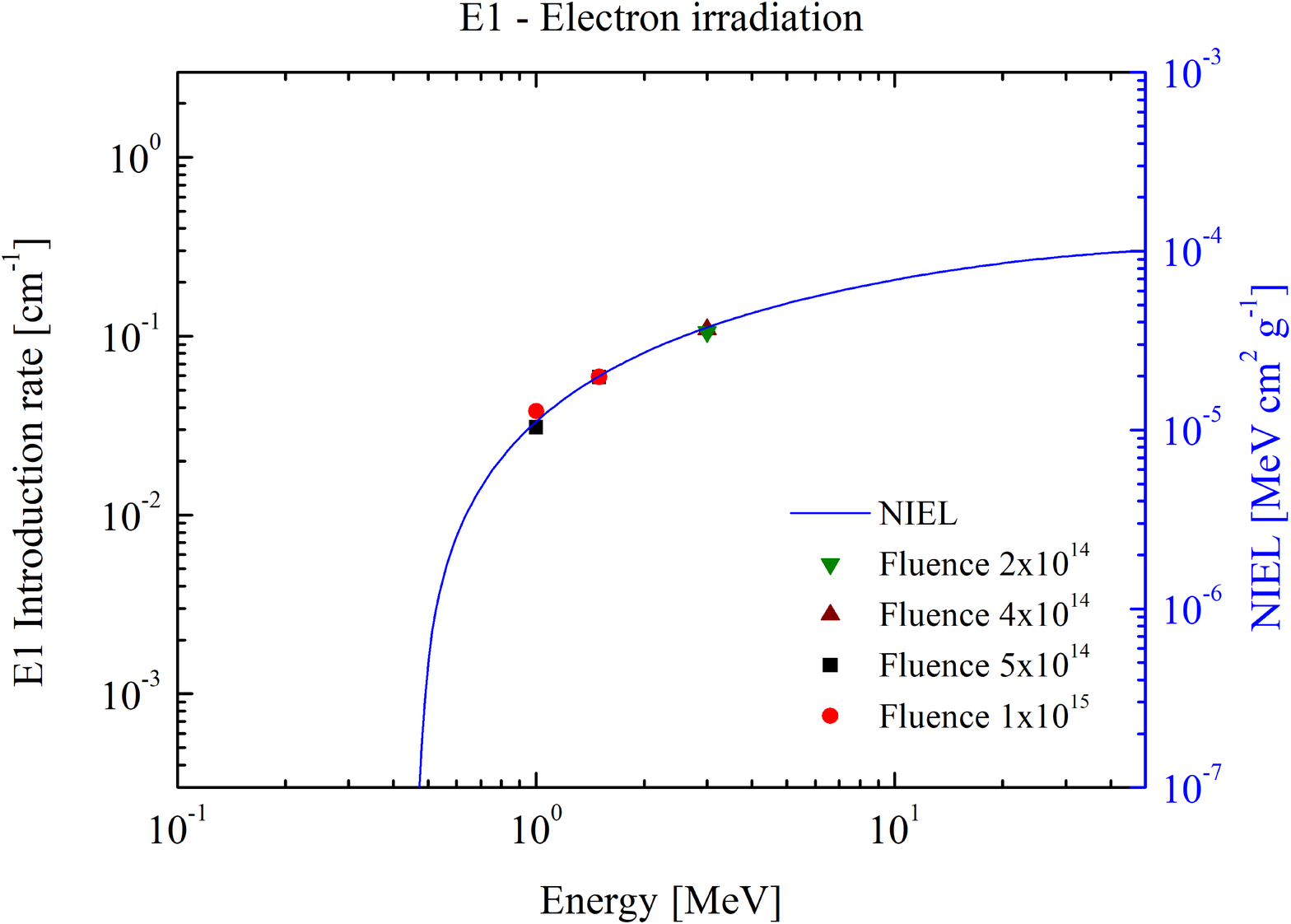}
\caption{E1 traps introduction rate as a function of incoming electron energy: right scale NIEL values in GaAs sub cell for electrons.}
\label{fig:10}
\end{figure}
\begin{figure}[h!]
\centering
\includegraphics[width=0.5\textwidth]{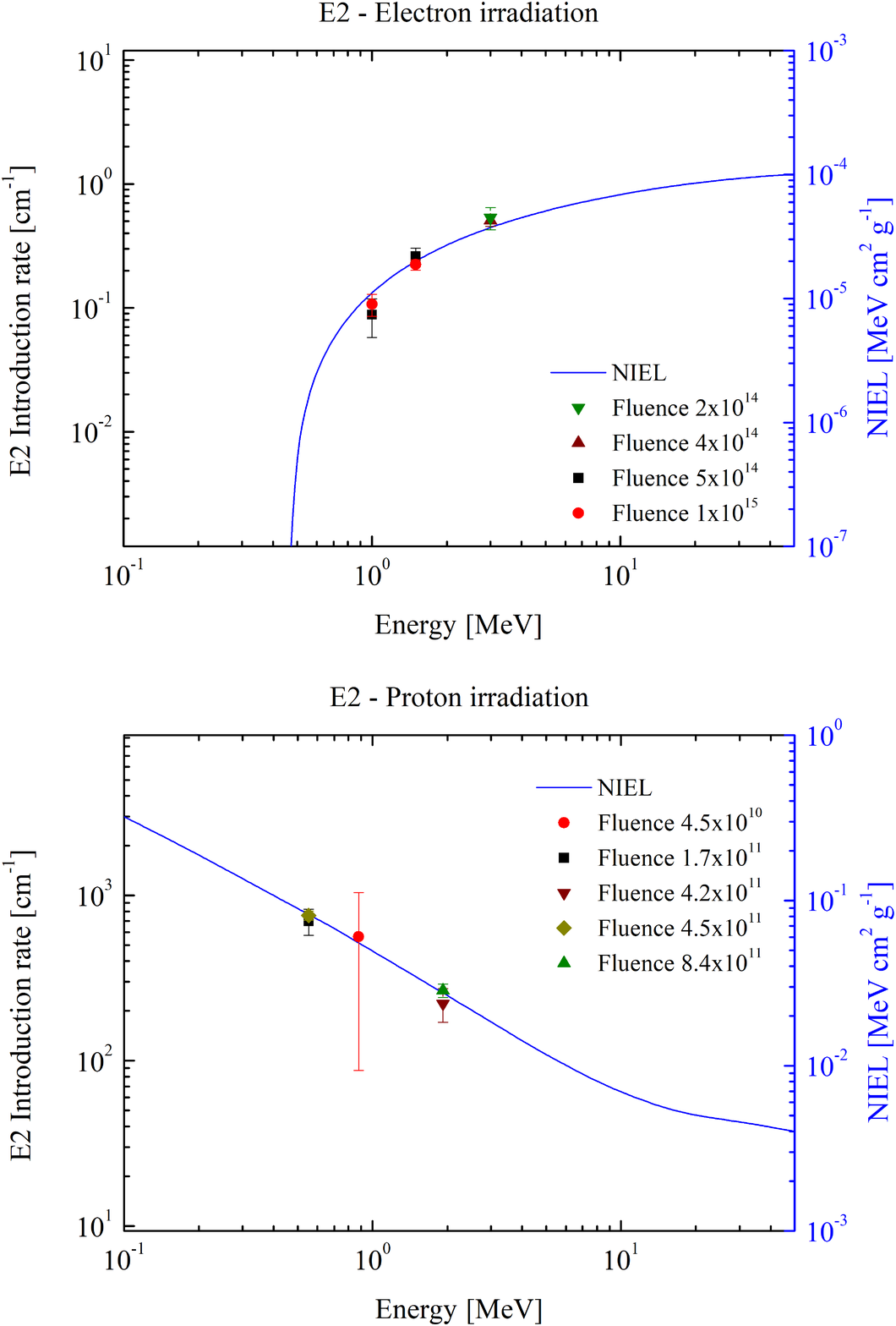}
\caption{E2 traps introduction rate as a function of incoming electron (top) and proton (bottom) energy: right scale NIEL values in GaAs sub cell for protons and electrons.}
\label{fig:11}
\end{figure}
By using these conversion factors, it can be seen from the Figures \ref{fig:10} and \ref{fig:11} that the NIEL curves well represent the introduction rates of the traps as a function of the energy E of the incident particle.

\section{Conclusion}
Triple junction (InGaP/GaAs/Ge) solar cells and related single junction (SJ) isotype sub-cells were irradiated with protons and electrons at different energies.\\
The data were analyzed by means of the Displacement Damage Dose (DDD) method based on the NIEL hypothesis.\\
It is demonstrated that by choosing an appropriate threshold energy, $E_{d}$, for atomic displacement a single degradation curve for each solar cell type and for each electrical parameter can be found.\\
To better understand the nature of the defects produced by irradiation, DLTS analysis is conducted on the real sub-cells manufactured as diodes and not on ``ad hoc''  p/n junctions typically used for this kind of measurements.\\
The DLTS analysis, carried out on middle junctions, indicates two main traps are introduced, E1 having an energy of 0.21 eV and E2 having an energy of 0.45 eV above the top of the valence band. These traps are for majority carriers because the high doping associated with the middle cell junction prevented the detection of minority carrier traps.\\
From the DLTS measurements it has been possible to correlate the defects concentration and the calculated Displacement Damage Doses. This analysis reveals that complex defects are likely to be introduced at a slightly different rate for electron and proton irradiations.\\
The set of measurements performed on this work experimentally support the validity of SR-NIEL treatment approach for obtaining the displacement damage doses.

\vspace*{0.5cm}
{\bf Acknowledgment:}{Part of this work was supported by ESA contract 4000116146/16/NL/HK with title "Non-Ionizing Energy Loss (NIEL) Calculation and Verification".}

\end{document}